 \def\gtorder{\mathrel{\raise.3ex\hbox{$>$}\mkern-14mu
	      \lower0.6ex\hbox{$\sim$}}}
 \def\ltorder{\mathrel{\raise.3ex\hbox{$<$}\mkern-14mu
	      \lower0.6ex\hbox{$\sim$}}}

 \def\rot{{\rm rot}}
 \def\min{{\rm min}}
 \def\max{{\rm max}}
 \def\ln{{\rm ln}}
 \def\m{{\rm Macho}\ }
 \def\dol{{D_{\rm OL}}}

 \def\dlmc{{D_{\rm LMC}}}
 \def\max{{\rm max}}
 
 \def\kms{{\rm km}\,{\rm s}^{-1}}
 \def\M{{\rm Macho}\ }
 \def\Mk{{\rm Macho}}
 \def\Ms{{\rm Machos}\ }
 \def\Msk{{\rm Machos}}

 \documentstyle[12pt]{article}
\begin{document}
 \large
 \centerline
 {\bf A Spectroscopic Method to Measure}
 \centerline{\bf  Macho Proper Motions
}

 \normalsize
 \vskip 2cm

 \centerline{\bf  Dan Maoz}

 \vskip .3cm
 \centerline{School of Physics \& Astronomy and Wise Observatory}

 \centerline{ Tel-Aviv University, Tel-Aviv 69978, ISRAEL}

 \centerline{E-mail: dani@wise.tau.ac.il}
 \vskip .3cm
 \centerline{and}
 \vskip .3cm
 \centerline{\bf Andrew Gould}
 \vskip .3cm
 \centerline{Department of Astronomy}
 \centerline{ Ohio State University, Columbus, OH 43210}

 \centerline{E-Mail: gould@payne.mps.ohio-state.edu}

 \vskip 4cm
 \centerline{submitted to {\it The Astrophysical Journal Letters}:
  January 3, 1994}
 \newpage
 \baselineskip=22pt

 \section*{Abstract}
A Massive Compact Halo Object  (Macho) that lenses a background star will
magnify different parts of the rotating stellar disk by varying amounts. The
differential magnification will cause a shift in the centroid of the star's
spectral lines during the lensing event. The shift is proportional to the
ratio of the stellar radius to the projected separation of the Macho from the
star. It therefore  provides a  direct measure of the  Einstein ring radius,
and so also a measure of the Macho's proper motion (angular speed). This
measurement can remove some of the degeneracy between mass, distance to the
lens, and transverse velocity that exists in the interpretation of results
from ongoing microlensing experiments, and is an independent test of the
lensing nature of the event. We show that  using the high precision attainable
by  stellar radial velocity measurements, it is possible to measure proper
motions for $\sim 10\%$ of Machos that lens A-stars in the Large Magellanic
Cloud (LMC), i.e.\  $\sim 7\%$ of the type of relatively high-magnification
events that have been reported to date. If this proper-motion measurement were
combined with a parallax measurement of the ``reduced velocity'', then the
Macho mass, distance, speed, and direction could each be separately
determined. The shift can be measured for $\sim 20\%$ of the  A-star events
generated by Machos in the dark halo of the LMC. This in turn would provide a
measurement of the fraction of LMC vs. Galactic Macho events.

 \bigskip
 Subject Headings: spectroscopy -- dark matter -- gravitational lensing --
 Magellanic Clouds
 \vfill
 \eject

 \section{Introduction}

	 Two independent groups have recently found a total of three
 candidate
 microlensing events, apparently caused by Massive Compact Halo Objects
 (\Ms) along the
 line of sight toward the Large Magellanic Cloud (LMC) (Alcock et al.\ 1993;
 Aubourg et al.\ 1993).  The events
 are achromatic, have maximum magnifications of $A_\max = 6.8$, 3.3, and 2.5,
 and characteristic times $\omega^{-1} = 17$ days, 30 days and 26 days.
 The light curves fit the  form first predicted by Paczy\'nski (1986)
 quite well:
\begin{equation}
 A[x(t)] = {x^2 + 2\over x(x^2 + 4)^{1/2}},\qquad
 x(t) = \sqrt{\beta^2 + \omega^2(t-t_0)^2}.
\end{equation}
 Here $t_0$ is the midpoint of the event, $\omega^{-1}$ is
 the characteristic time, and $\beta$ is the dimensionless
 impact parameter (normalized to the Einstein ring radius).

	 Of the three measurable parameters in equation (1), only the
 characteristic time, $\omega^{-1}$ yields any information about the \Msk.
 It is related to the underlying physical parameters by
\begin{equation}
 \omega^{-1} = {\theta_*\dol\over v}=
 {[4GM \dol(1-\dol/\dlmc)]^{1/2}\over v c},
\end{equation}
 where $\theta_*$ is the Einstein ring radius,
 $M$ is the \M mass, $\bf v$ is its transverse velocity, and
 $\dol$ is its distance from the observer.  The distance of the lensed star is
 denoted $\dlmc$.  For any given event, one cannot determine the mass,
distance,
 and speed separately.  For an ensemble of events, the typical speed and
 distance are expected to be $v\sim 200\, \kms$ and $\dol\sim 10$ kpc,
 respectively.  Hence, $M\sim (70 \omega\, {\rm days})^{-2} M_\odot$.
 If the first few events prove typical, it will imply a mass scale of
 $\sim 0.1\, M_\odot$, near the hydrogen-burning limit.

	 One would like to determine as much as possible about the distribution
 of \Msk, not just the mass scale.  For example, one would like to know if
 the \Ms  actually lie in a halo as opposed to a disk or thick disk,
 whether the halo is spherical or oblate, whether the  halo is
 truncated or extends all the way to the LMC, whether it is rotating, whether
 \Ms populate the LMC as well as the Galaxy, and what the distribution of
 \M masses is.
	 To extract these additional pieces of information, new methods of
 analyzing the data as well as new experiments are required.

 	Gould  (1992, 1993, 1994a,b), Sackett \& Gould (1993), and
 Gould, Miralda-Escud\'e \& Bahcall
 (1994) have developed a number of different techniques
 for extracting additional parameters from
 lensing events.
 In particular, Gould (1994a) showed that the non-zero size of the lensed
 stellar disk modifies the light curve of the event in a manner that allows
 one to measure the \Mk's Einstein radius, $\theta_*$, and hence its proper
 motion (angular speed), $\omega\theta_*$.
In practice, however,
 the effect is measurable only for a fraction of
 events during  which the \M passes  over the line of
 sight to the face of the source star,
  $(1/550)(M/0.1\,M_\odot)^{-1/2}$ for \Ms in the Galactic halo
 and $(1/70)(M/0.1\,M_\odot)^{-1/2}$ for \Ms in the LMC.

 In this {\it Letter}, we predict
 a second, spectroscopic, effect  which
also arises from the non-zero
 size of the source, combined with the source rotation.
 Differential magnification of the stellar disk during the event  induces a
 shift in the stellar spectral lines.  By measuring the shift
 as a function of time one can determine the Einstein ring radius and so the
 proper motion of the \Mk.
 In contrast to the photometric effect analyzed by Gould (1994a),
 this spectroscopic effect can be measured even when the \M is many stellar
 radii from the source.  Hence, the proper-motion measurement can be made
 for a larger fraction of events.
 We estimate that with  an 8m telescope the effect can be measured in $\sim
 7\%$ of photometrically detected relatively high-magnification  ($\beta
\ltorder 0.5$) \M events.
 The effect can be measured in $\sim 15\%$ of high-magnification
events  generated by
\Ms in the LMC.

	Measurement of the proper motion of Galactic \Ms would, in
itself, remove some of the degeneracy among the three \M parameters, mass,
distance, and transverse speed.  However, if this
measurement were combined with a parallax measurement of the ``reduced
velocity'', then the degeneracy could be completely broken and the three
parameters plus the transverse direction could be
separately determined.

	The proper motion of Galactic \Ms is $\sim 15$ times greater than
that of LMC \Msk.  Hence, a proper-motion measurement clearly distinguishes
between the two.
By identifying even a handful of LMC \Msk, one could
determine the fraction of events generated by them.  This would both provide
direct information about the LMC halo and remove a major background to
the primary Galactic signal (see Sackett \& Gould 1993).

 \section{Spectral Shift in Lines of a Microlensed Star}

 The spectral line profile of an unlensed star is broadened by the
 effect of the star's rotation.
 Different parts of a finite stellar disk
 are magnified by varying amounts during a \M lensing event.
  In order to determine the time-dependent rotational line profile, we apply
 equation (1) to calculate the magnification at each point on the star.
 Expressing all angles in units of the Einstein radius, $\theta_*$, the line
profile is
 \begin{equation}
 f(v,t) = C\int_{-\sqrt{\rho^2-z_1^2}}^{\sqrt{\rho^2-z_1^2}} d\,z_2
 A(|{\bf x}(t) + {\bf z}|)S({\bf z}),\qquad v \equiv {z_1\over \rho} v_\rot
 + v_0
 \end{equation}
 where $\rho$ is the stellar radius, ${\bf x}(t)$ is the projected separation
 of the stellar center from the \Mk, $S({\bf z})$ is the surface brightness
 as a function of position, $\bf z$, on the stellar disk, $v_\rot$ is the
 projected rotation speed of the star, $v_0$ is the velocity of the stellar
 center of mass, and $C$ is a constant.  The $z_2$ axis is defined to be
 aligned with the stellar rotation axis.
The effect of limb-darkening
can be approximated
(e.g. Allen 1973) by the form:
\begin{equation}
S({\bf z})
=1-u_2-v_2+u_2(1-|{\bf z}|^2/\rho^2)^{1/2}+v_2(1-|{\bf z}|^2/\rho^2),
\end{equation}
with $u_2=0.99$ and $v_2=-0.17$ at 4500\AA. We assume the star is devoid
of star-spots.

	The magnitude of the separation function $x(t)= \sqrt{\beta^2 +
 (\omega t)^2}$ is known from the overall light curve.
 One can integrate equation (3) numerically and
 compare the expected profiles with the rotational profiles observed in a
series of
 measurements.
 It will generally be more practical to measure the shift in the {\it centroid}
 of the
 line, i.e. the star's apparent radial velocity, as a function of time:
 \begin{equation}
\langle v(t)\rangle =\frac{\int dv f(v,t)v}{\int dv f(v,t)}.
 \end{equation}
  Apart from the rotational broadening described by equation (3), stellar line
profiles
can be intrinsically broadened by thermal and turbulent motions, by the natural
line width, and by pressure broadening.
The intrinsic broadening does not, however, affect the line centroid.

To study the behavior of the velocity shift, we  first make the simplifying
assumptions that
 $\rho\ll x \ll 1$
 and $S({\bf z})=\,$constant.
The integral for the line profile (eq.\ 3)  can then be approximated as
 \begin{equation}
 f(v,t) = C{\rho\over v_\rot}\sqrt{v_\rot^2 - (v-v_0)^2}
 \biggl[1 + {v-v_0\over v_\rot}\,\rho\,
{\beta\cos\alpha+\omega t \sin\alpha\over \beta^2+(\omega t)^2}\biggr],
 \end{equation}
and the line centroid (eq.\ 5) becomes
 \begin{equation}
 \langle v(t)\rangle
 = v_0 + {\xi\over 4} v_\rot\,\rho\,
{\beta\cos\alpha+\omega t \sin\alpha\over \beta^2+(\omega t)^2},
 \end{equation}
where  $\xi=1$ and $\alpha$ is the angle
 between
the \M velocity $d {\bf x}/d t$
and the star's projected rotation axis.  For a limb-darkened stellar profile
as described by equation (4), equation (7) remains valid with
$\xi = (30 - 14 u_2 - 20 v_2)/(30 - 10 u_2 - 15 v_2)$,  e.g.,
$\xi=0.86$ at 4500\AA.
  We find numerically
that for $\rho\ltorder x\ll 1$, the velocity shift in equation (7) should
be reduced by a factor $[1-0.2(\rho/x)^2]$.

 Figure 1 shows the shift in the
 line centroid,  $\Delta\langle v(t)\rangle\equiv \langle v(t)
\rangle - v_0$
as a function of time for an event with timescale
$\omega ^{-1}=17$ days and a star with $v_{\rot}$=100 km s$^{-1}$.
 The dependence of the line-shift curve on the parameters
 $\rho$, $\beta$, and $\alpha$, the angle between the stellar
 axis and the \M
trajectory, is illustrated in the separate panels.
 As reference, for a  $0.1 M_{\odot}$ \M in the Galactic halo at a
distance of 10 kpc
  lensing an A star of radius $2.5R_{\odot}$ in the LMC,
$\rho=0.001$. Such a \M in the
 LMC halo, lensing an LMC star 2.5 kpc behind it, will have
$\rho=0.01$. The fraction of lensing
 events having dimensionless impact parameter of $\beta$ or less (and hence
maximum magnification $\beta^{-1}$ or greater) is simply $\beta$.
 All values of the angle $\alpha$ are equally likely.

 As can be seen from equation (7) and Figure 1, varying the value of
$\rho$, the ratio of the apparent
  stellar radius to the
Einstein radius, affects only the amplitude
 of the line-shift curve. Changes in the impact parameter $\beta$ affect
both the
 amplitude of the curve and its shift in time relative to the time of maximum
 total magnification. Changes in the projected angle $\alpha$ between the
stellar spin-axis
 and the \M trajectory affect the amplitude, the time-shift, and the shape
 of the curve. The impact parameter $\beta$ can be fit independently
 to the photometric light curve.
The parameters $\rho$ and $\alpha$ can then be fit unambiguously to an
 event, given enough  signal to noise, and the one interesting parameter,
$\rho$, measured.
Since the physical stellar radius is known from the star's spectral type, as is
the distance to the LMC, $\rho$ provides a measure of the \M Einstein
ring  radius, $\theta_*$
and the \M proper motion $\omega \theta_*$. Knowledge of the
proper motion removes one part of the
degeneracy between \M mass,
distance, and transverse velocity in the interpretation
of the photometric light
curve (see eq.\ 2). For example, for an assumed velocity $v$,
the distance $D_{OL}$
to the lens can be derived, and then from this distance and $\theta_*$,
the mass $M$  can be calculated.

 From equation 7  and
 the relevant range of parameters,  a velocity resolution
  of  ${ \delta
v}/{v_{\rot}}\sim \rho
\sim 0.01$--$ 0.001$
 is required in order to measure $\rho$ in some fraction of lensing events.
 A temporal resolution of order hours is needed.
 We examine the observational aspects more carefully below.

 The stars being monitored by the  MACHO and EROS collaborations are
 $\sim 1/3$
 main-sequence stars, mostly A stars,
and $\sim 2/3$ red giants (C.\ Alcock 1993, private communication).
The projected
 rotation speeds of A stars have a broad distribution with a mean
$v_{\rot}$=150 km s$^{-1}$
 for A0 types,
 decreasing to  $v_{\rot}$=80 km s$^{-1}$ for F0
 (e.g. Allen 1973).
Giants bluer than about G2 have similar rotation properties, with mean
  $v_{\rot}\sim 75$ -- 55 km s$^{-1}$
for F0 III to G0 III. A ``rotation boundary'' exists
for giants at spectral type G3
(Gray 1989; a similar boundary exists in the main sequence
at F5). Giants redder than G3 rotate very slowly, with mean velocities of
 $v_{\rot}\sim 5$ -- 2 km s$^{-1}$
for G3 III to K3 III.
Giants later than G3 rarely or never
rotate faster than 9 km s$^{-1}$.
Furthermore, their line broadening is dominated
by photospheric macroturbulence of
order   $\sim 4$ -- 8 km s$^{-1}$ (Gray 1989), which affects the accuracy
with which their $v_{\rm rot}$ can be determined.  (Below, in \S\ 3, we will
show that the accuracy of the proper-motion measurement is fundamentally
limited by how well $v_{\rm rot}$ is known.)\ \
  A precision $\delta v\sim 100$ m s$^{-1}$ is required for A stars
and $\sim 5$ m s$^{-1}$ for red giants.

 Such precisions can be obtained today in stellar radial-velocity
 measurements. For example, the stellar radial-velocity survey
 at the CfA achieves accuracies $\sim 1$ km s$^{-1}$ for 12 mag A stars
 and $\sim 200$ m s$^{-1}$ for G-stars in 20-minute exposures
 on a 1.5m telescope (Latham 1992; T.\ Mazeh  1993, private communication).
The survey uses an
intensified reticon detector, which can utilize only
one echelle order. Since the survey is not optimized for A stars, the
one order includes only weak metal lines, and none of the
Balmer lines which are strong in A stars.    Modern CCD
detectors can have at least 5 times higher quantum efficiency.
CCDs can simultaneously
record many echelle orders and so increase the observed number of lines and
hence the
precision.
The precision of radial velocity measurements increases as
 (No.photons$\times$No.lines)$^{1/2}$, but is ultimately
limited by systematic effects, such as spectrograph stability.
It is plausible that
 a 17 mag A star
 in the LMC could be measured to
$\sim 100$ m s$^{-1}$ in a one-hour exposure with a CCD on an 8m telescope with
a spectroscopic setup optimized for A stars.
(Recall that the stars being monitored, with
 typical unlensed magnitudes of 19-20, are in the process of being magnified by
 several magnitudes.)\ \
 While improved velocity resolution is possible for red giants because of
their numerous
 and strong lines, techniques yielding resolutions of $<10 $ m s$^{-1}$
have been
 applied only to very bright ($V\sim 5$ mag) stars (e.g. Walker 1992),
  and  may not be feasible here.
Because of this difficulty as well as the previously mentioned problem of
determining the rotation speed of red giants, it may not be possible to
make a spectroscopic determination of the proper motion of \Ms which lens
this class of star.  When we estimate the fraction of measurable events
below, we therefore consider only A stars.

\section{ Error Estimates}

We examine now the errors that would be involved in an actual measurement.
Parameterizing the separation
$x$ by an angular
variable $\theta$,
\begin{equation}
x\equiv \beta\sec\theta,
\end{equation}
 equation (7) becomes
 \begin{equation}
\langle v\rangle = v_0 + {v_\rot\xi\over 4 \beta} \rho\cos\alpha\cos^2\theta +
 {v_\rot\xi\over 4 \beta} \rho\sin\alpha\cos\theta\sin\theta =
\sum_{i=0}^2 a_i f_i(\theta).
\end{equation}
 where
\begin{equation}
a_0\equiv {v_0\over Q}, \qquad a_1 \equiv\rho\sin\alpha,
 \qquad a_2\equiv\rho\cos\alpha,
\end{equation}
\begin{equation}
f_0(\theta)\equiv Q,\qquad f_1(\theta)\equiv Q\sin\theta\cos\theta,
 \qquad f_2(\theta)\equiv Q\cos^2\theta,
\end{equation}
 and $ Q\equiv v_\rot\xi/4\beta$. (Note that by including $Q$ in the
trial functions, $f_i(\theta)$, we have implicitly assumed that $Q$ is
known.  If there is a fractional uncertainty in $Q$ -- due for example
to an uncertainty in $v_{\rm rot}$ -- this will induce a similar uncertainty
in $\rho$ over and above those analyzed below.)\ \

	 We now suppose that a series of measurements are made of the velocity
 centroid and that the errors in these measurements scale inversely
 as the square root of the product of the exposure time and
 the luminosity.
 We normalize these to the error of a one hour observation at maximum
 magnification, which we take to be $\sigma_0 = \delta v$.  Hence for a
 general observation of length $\delta t$ and position $\theta$,
\begin{equation}
\sigma^2(\theta) = (\delta v)^2{{\rm hr}\over \delta t}\sec\theta.
\end{equation}
 We note that during the time interval $d t$, the lens moves by
$d\theta = \cos^2\theta \omega/ \beta d t$,
 where $\omega^{-1}$ is the time required to cross an Einstein ring radius.
 According to standard linear theory (Press et al.\ 1989), the inverse
 covariance matrix $b_{ij}$ for the three parameters $a_i$ is given by
\begin{equation}
 b_{ij} = \sum_k {f_i(\theta_k)f_j(\theta_k)\over \sigma^2(\theta_k)}.
\end{equation}
 For purposes of illustration, we assume that the observations are conducted
 continuously between $\theta_\min$ and $\theta_\max$.  Then, the above
 sum can be converted into an integral
\begin{equation}
 b_{ij} = \Delta^{-2}\int_{\theta_\min}^{\theta_\max} d\theta
 \left(\matrix{\sec\theta & \sin\theta & \cos\theta\cr
	       \sin\theta & \sin^2\theta\cos\theta & \sin\theta\cos^2\theta\cr
	       \cos\theta & \sin\theta\cos^2\theta & \cos^3\theta}\right),
\end{equation}
where
 \begin{equation}\Delta \equiv {4\delta v\over
v_\rot\xi}\sqrt{\beta\omega\rm hr}.\end{equation}
 For example, for an event with peak magnification of 5,
 $\omega=(17{\rm days})^{-1}$, and 8 hours observations per day,
$\Delta ={4\delta v /
(v_\rot\xi)}\sqrt{0.2 (17)^{-1}/ 8}$.

	 For simplicity, we consider the case where the observations are
 made symmetrically around the peak of the light curve, $\theta_\max
 =-\theta_\min=\theta_0$.  The total length of the observations is then
 $T = 2(\beta/\omega)\tan\theta_0$.
 For this case, the even and odd components
 decouple and $b_{ij}$ takes a relatively simple form.  We invert $b$ to
 find the covariance matrix $c\equiv b^{-1}$.  The relevant terms are
 \begin{equation}
c_{11} = {3\over 2}\csc^3\theta_0\Delta^2,\qquad
 c_{22} = {1\over 2}\biggl(\sin\theta_0 - {\sin^3\theta_0\over 3}
 - {\sin^2\theta_0\over \ln\,\tan(\theta_0/2+\pi/4)}\biggr)^{-1}\Delta^2,
\end{equation}
 for the variances of
 $a_1\equiv \rho\sin\alpha$ and $a_2\equiv\rho\cos\alpha$, respectively.
 Note that the quantity of interest is $\rho=\sqrt{a_1^2+a_2^2}$.

	In the above formal treatment, we have implicitly assumed that
the source star is moving with constant velocity relative to the Earth.
However, in general there will be some relative acceleration due to the motion
of the Earth about the Sun and to the possible motion of the source about
an unseen companion.  The Earth's motion
is known and can be
taken out.  The motion of the source could likewise be taken out if it were
measured by long-term observation of the source radial velocity.  However,
it is much easier to measure the source velocity when  the source
 is highly magnified,
so the best constraint on the acceleration of the source may come from the
observations during the lensing event.  This acceleration will be very
nearly linear provided that the orbital period is much larger than the span of
observations.  We assume that this is the case.

The effect of an acceleration, $g$,
is to add a term which is linear in time, i.e. $\propto \tan\theta$.
Formally we write
$f_3(\theta) = Q\tan\theta;\quad a_3 = {g/ Q}$.
Once this additional degree of freedom is allowed, the covariance
matrix becomes four-dimensional.  However, under the assumption of a
symmetric interval of observations, the even and odd components decouple.
We can therefore examine two smaller submatrices, $b^+$ for components
0 and 2, and $b^-$ for components 1 and 3.
For the odd terms,  we find
\begin{equation}
b^-_{ij} = \Delta^{-2}\int_{-\theta_0}^{\theta_0} d\theta
\left(\matrix{\sin^2\theta\cos\theta & \sin\theta\tan\theta\cr
              \sin\theta\tan\theta & \tan^2\theta\sec\theta}\right),
\end{equation}
which implies a variance of  $a_1=\rho\sin\alpha$ of
\begin{equation}c^-_{11} = {1\over 2}\biggl({\sin^3\theta_0\over 3} -
2{[\ln\,\tan(\theta_0/2+\pi/4)-\sin\theta_0]^2\over \tan\theta_0\sec\theta_0
-\ln\,\tan(\theta_0/2+\pi/4)}\biggr)^{-1}\Delta^2.\end{equation}
The variance $c^+_{22}$ of  $a_2=\rho\cos\alpha$
remains as in equation (16). The  rms uncertainty in $\rho$ (averaged over
all angles $\alpha$) is
$\delta \rho = [(c_{11}+c_{22})/2]^{1/2}.$

Figure 2 shows  $\delta \rho / \rho$ as a function of the total time span
of the observations $T$, where we have assumed $\omega^{-1}=17$ days and
$\delta v/ v_{\rm rot}\xi
=10^{-3}$ per one-hour observation at maximum magnification.
The two panels show the errors
in $\rho$ for $\rho=0.001$ (typical Galactic \Msk)
and for $\rho=0.01$ (typical \Ms
in the LMC) for various values of the impact parameter $\beta$.
 We have assumed 8 hours of observation per day in the
 Galactic \M panel ($\rho=0.001$),
and just one hour per day for the LMC halo \Ms (or, equivalently, a longer
observation on a smaller telescope, achieving the same $\delta v/v_{\rm rot}$).
  Assuming
one is interested in $<30\%$ error in $\rho$  (i.e., a $3\,\sigma$
detection) and one can dedicate
$<8\,$days
of telescope time to the monitoring of the \M event, we see that $\beta< 0.1 $
is required for typical Galactic events.
 One measurement per night for 8 days would be sufficient to detect
LMC \Ms with $\beta<0.2$.
If the LMC has a dark halo, $\sim 10\%$ of all events are
expected to be LMC events
(Gould 1993). Hence $\sim 10\%$ of all A-star events can be
detected at the $3\,\sigma$ level.
Note that the detection
capabilities of the MACHO and EROS experiments
are biased toward  low-$\beta$ events, and that all their
detected events so far
have magnification $>2$ (i.e. $\beta < 0.5$). The velocity shift is therefore
measurable in $\sim 20\%$ of the A-star events of the type that is
being detected,
or $\sim7\%$ of the events being detected on all stars.

\section{Discussion and Conclusions}
We predict the existence of a spectroscopic line-shift, resulting
from differential magnification of a rotating stellar disk, during microlensing
events in the ongoing MACHO and EROS experiments. The effect can be
measured and monitored on large telescopes for a fraction of the A-star
lensing events
(A-stars
should supply $\sim 1/3$ of the events), provided that the spectroscopic
observations can be initiated fast enough
to catch the event before maximum light. The spectroscopic signature
can serve as an additional test of the lensing (as opposed to variable-star)
nature of the event. More importantly, the Einstein ring  radius
can be measured
from the temporal behavior of the line-shift.
This technique can therefore remove part of the degeneracy in the
derivation of \M parameters in the ongoing experiments.

The spectral
shift can be measured with relatively smaller effort (one measurement
per night) in a  large fraction
of microlensing events involving \Ms in the halo of the LMC itself, which
can then be recognized as such. If, for example, 10\% of the events are
due to LMC \Msk,
 spectroscopic measurements of 50 A-star events
 with  $\beta < 0.2$ for one hour per night for a week will reveal
the effect in  $\sim 5$ cases, and
 would provide a direct measurement of the LMC
fraction of \M events.  The LMC events can be distinguished from any Galactic
``foreground'' events by their much lower proper motion (Gould 1994a).

 With a more ambitious program in which events are followed for a week
with eight measurements per night (likely requiring an 8m telescope),
the proper motion can also be measured for the
A-star
+ Galactic \M events with
$\beta\ltorder 0.10$.
 While this is a lot of telescope time the potential payoff is very
great, particularly if the ``reduced velocity'' can be measured from
parallax observations (Gould 1992, 1994b; Gould et al.\ 1994).  The
reduced speed is given by $(D_{\rm OL}^{-1}-D_{\rm LMC}^{-1})^{-1}
\omega\theta_*$.  Hence, by measuring the proper motion, $\omega\theta_*$
one would find the distance to the \Mk, $\dol$.  The distance and the
Einstein radius, $\theta_*$, yield the mass, $M$.  Combining these with
$\omega$ gives the transverse speed.  In general, measurement of the
reduced velocity requires observation from two small satellites in
solar orbit.  It is likely that the reduced speed could be obtained
from only one such satellite (Gould 1994b).  Gould et al.\ (1994)
show that it is generally
possible to measure one component of the reduced velocity
from ground-based observations provided that the \Ms are going of order
the Earth's orbital speed of $30\,\rm km\,s^{-1}$.  This would be
extremely rare for \Ms in the halo.  However, the measurement proposed
by Gould et al.\ actually becomes much more sensitive for very
high-magnification events  $(\beta\ltorder 0.1)$
such as those for which proper-motion determinations
can be made.  For these events, we conjecture that one or perhaps
even both components of the
transverse velocity could be measured from the ground, even for \Ms traveling
at halo speeds, $\sim 200\,\rm km\,s^{-1}$.

{\bf Acknowledgements}: We would like to thank D.\ Goldberg, T.\ Mazeh,
and C.\ Pryor for helpful discussions.  D.M. acknowledges support through
an Alon Fellowship.

\newpage
\section*{References}

\renewcommand\bibitem{\par\noindent\hangindent=2pc\hangafter=1}

\bibitem{Alcock, C., et al.\ 1993, Nature, 365, 621}
\bibitem{Allen, C.W. 1973, { Astrophysical Quantities} (London: Athlone)}
\bibitem{Aubourg, E., et al.\ 1993, Nature, 365, 623}
\bibitem{Gould, A.\ 1992, ApJ, 392, 442}
\bibitem{Gould, A.\ 1993, ApJ, 404, 451}
\bibitem{Gould, A.\ 1994a, ApJ Letters, 421, in press}
\bibitem{Gould, A.\ 1994b, ApJ Letters, 421, in press}
\bibitem{Gould, A., Miralda-Escud\'e, J.\ \& Bahcall, J.\ N.\ 1994,
ApJ Letters, in press}
\bibitem{Gray, D.F. 1989, ApJ, 347, 1021}
\bibitem{Latham, D.W. 1992, in {Complementary Approaches to
Double and Multiple Star Research--ASP Conference Series Vol.\ 32},
eds.\ H.A.\ McAlister and W.I.\ Hartkopf, (San Francisco: ASP), p.\ 110}
\bibitem{Paczy\'nski, B.\ 1986, ApJ, 304, 1}
\bibitem{Press, W.\ H., Flannery, B.\ P., Teukolsky, S.\ A., \&
Vetterling, W.\ T.\ 1989, Numerical Recipes (Cambridge Univ.\ Press)}
\bibitem{Sackett, P.\ D.\ \& Gould, A.\  1993, ApJ, 419, 648}
\bibitem{Walker, G.A.H. 1992, in {Complementary Approaches to
Double and Multiple Star Research--ASP Conference Series Vol.\ 32},
eds. H.A.\ McAlister and W.I.\ Hartkopf, (San Francisco: ASP), p.\ 67}
\vfill
\eject
\section*{Figure Captions}
{\bf Figure 1} The line centroid shift $\Delta v$ in a microlensed star
as a function of time. The plots are for an Einstein-radius crossing
time $\omega^{-1}=17$ days, and a projected stellar rotation speed
$v_{\rot}=100$ km s $^{-1}$. (The horizontal and vertical axes can
be scaled for any $\omega^{-1}$ and $v_{\rot}$). The separate panels
show the dependence of the velocity curve on the three parameters:
 dependence
(a) on $\rho$, the ratio of the stellar angular radius to
 the Einstein radius; (b) on $\beta$, the ratio
of the impact parameter to the Einstein radius; (c) on
$\alpha$, the angle between the \M velocity vector
and the projected stellar rotation axis.
$\alpha$ can be deduced from the shape of the
velocity curve, and $\beta$ from the total photometric magnification
of the event. The amplitude of the velocity curve then yields
$\rho$, or, since the stellar radius is known, the Einstein radius
of the \Mk.

{\bf Figure 2} The expected relative error in $\rho$,
 the ratio of the  stellar angular radius to
 the Einstein radius, as a function of the time span of the
observations, $T$ in days. The two panels are for $\rho=0.001$ (typical
for $0.1 M_{\odot}$ \Ms in the halo of the Galaxy) and for  $\rho=0.01$
(typical for \Ms of this mass in the LMC halo). The various curves
give the dependence of the error on the dimensionless impact parameter
$\beta$. Since a fraction $\beta$ of lensing events have impact parameter
$<\beta$, the fraction of lensing events where  $\delta\rho/\rho$ can
be measured to a specified accuracy given a stretch of telescope time $T$
can be read off the plots. It is assumed that the timescale of the event
is  $\omega^{-1}=17$ days and the precision of the radial velocity measurement
that can be obtained in 1 hour at the time of maximum magnification is
 $\delta v/ v_{\rm rot}=10^{-3}$.
Eight hours of observation per day are assumed in the $\rho=0.001$ panel
and just one hour (or one measurement with equivalent precision)
per day
in the $\rho=0.01$ panel. About 10\%
of Galactic \M events and 20\% of LMC \M events involving A stars
can be measured to better than 30\% accuracy in 8 days of observing
per event.

\end{document}